\documentclass[prl,twocolumn,twoside,preprintnumbers,superscriptaddress,nofootinbib]{revtex4}
\usepackage{amsmath,slashed}
\usepackage{graphicx,graphics}
\usepackage{dcolumn}
\usepackage[hyperfootnotes=false]{hyperref}
\usepackage{xspace}
\usepackage{color}

\newcommand{\mpi}{M_{\pi}}
\newcommand{\Order}{\mathcal{O}}

\newcommand{\GeV}{\,\text{GeV}}
\newcommand{\mk}{M_K}
\newcommand{\mtau}{m_\tau}

\newcommand{\pk}{p_K}
\newcommand{\ppi}{p_\pi}
\newcommand{\pnu}{p_\nu}
\newcommand{\diff}{\text{d}}

\newcommand{\Lagr}{\mathcal{L}}
\newcommand{\beq}{\begin{equation}}
\newcommand{\eeq}{\end{equation}}

\renewcommand{\Im}{\text{Im}\,}

\begin{document}

\preprint{INT-PUB-17-055, LA-UR-17-31337, PSI-PR-17-21}

\title{A no-go theorem for non-standard explanations of the $\boldsymbol{\tau\to K_S\pi\nu_\tau}$ $\boldsymbol{CP}$ asymmetry}

\author{Vincenzo Cirigliano}
\affiliation{Theoretical Division, Los Alamos National Laboratory, Los Alamos, NM 87545, USA}
\author{Andreas Crivellin}
\affiliation{Paul Scherrer Institut, PSI, CH--5232 Villigen, Switzerland}
\author{Martin Hoferichter}
\affiliation{Institute for Nuclear Theory, University of Washington, Seattle, WA 98195-1550, USA}

\begin{abstract}
The $CP$ asymmetry in $\tau\to K_S\pi\nu_\tau$, as measured by the BaBar collaboration, differs from the Standard Model prediction by $2.8\sigma$. 
Most non-standard interactions do not allow for the required strong phase 
needed to produce a non-vanishing $CP$ asymmetry, 
leaving only new tensor interactions as a possible mechanism. 
We demonstrate that, contrary to previous assumptions in the literature, the crucial interference between vector and tensor phases is suppressed by at least two orders
of magnitude due to Watson's final-state-interaction theorem. 
Furthermore, we find that the strength of the relevant $CP$-violating tensor interaction 
is strongly constrained by bounds from the neutron electric dipole moment and $D$--$\bar{D}$ mixing.  
These observations together imply that it is extremely difficult to explain the current $\tau\to K_S\pi\nu_\tau$ measurement in terms of physics beyond the Standard Model originating in the ultraviolet.
\end{abstract}

\maketitle

\section{Introduction}

The presence of the baryon asymmetry in the universe is one clear indication that there has to be physics beyond the Standard Model (SM) of particle physics~\cite{Sakharov:1967dj}, since $CP$ violation within the SM, originating solely from the CKM matrix~\cite{Kobayashi:1973fv}, is far too small to explain the observed asymmetry~\cite{Cohen:1993nk,Riotto:1999yt}.
This need for additional $CP$ violation renders $CP$-violating observables particularly interesting probes of beyond-the-SM (BSM) physics, with potentially profound implications for the SM and the physics of the early universe.

$CP$ violation was first observed in the neutral kaon system~\cite{Christenson:1964fg}. $K^0$ and $\bar K^0$ mix into the mass eigenstates $K_S$ and $K_L$, which decay predominantly into $2\pi$ and $3\pi$, respectively. However, $K^0$--$\bar K^0$ oscillations induce the $CP$-violating decays $K_L\to\pi\pi$ 
at a level of $\Order(10^{-3})$.  In addition to this indirect mechanism, the SM also permits direct $CP$ violation, suppressed by another three orders of magnitude compared to indirect $CP$ violation~\cite{AlaviHarati:2002ye,Batley:2002gn}.

The focus of this article is the $CP$ asymmetry in the decay width $\Gamma$ of $\tau\to K_S\pi\nu_\tau$
\beq
 A_{CP}^\tau=\frac{\Gamma(\tau^+\to\pi^+ K_S\bar \nu_\tau)-\Gamma(\tau^-\to\pi^- K_S \nu_\tau)}{\Gamma(\tau^+\to\pi^+ K_S\bar \nu_\tau)+\Gamma(\tau^-\to\pi^- K_S \nu_\tau)}.
\eeq
In the SM the dominant contribution again arises indirectly from $K^0$--$\bar K^0$ mixing~\cite{Bigi:2005ts}, and the same statement holds for the analogous decays of $D$ mesons~\cite{Lipkin:1999qz}  
\begin{align}
\label{D_CP}
 A_{CP}^D&=\frac{\Gamma(D^+\to\pi^+ K_S)-\Gamma(D^-\to\pi^- K_S)}{\Gamma(D^+\to\pi^+ K_S)+\Gamma(D^-\to\pi^- K_S)}\notag\\
 &=-4.1(9)\times 10^{-3},
\end{align}
where the experimental number refers to the average of~\cite{Link:2001zj,delAmoSanchez:2011zza,Ko:2012pe,Bonvicini:2013vxi}, see~\cite{Amhis:2016xyh}.
In fact, the amplitude governing the indirect $CP$ violation can be extracted very accurately from semileptonic kaon decays~\cite{Patrignani:2016xqp} 
\begin{align}
 A_L&=\frac{\Gamma(K_L\to\pi^-\ell^+\nu_\ell)-\Gamma(K_L\to\pi^+\ell^-\bar\nu_\ell)}{\Gamma(K_L\to\pi^-\ell^+\nu_\ell)+\Gamma(K_L\to\pi^+\ell^-\bar\nu_\ell)}\notag\\
 &=3.32(6)\times 10^{-3},
\end{align}
where $\ell=e,\mu$, and, neglecting small corrections from direct $CP$ violation, 
\beq
\label{CP_SM}
A_{CP}^{\tau,\text{SM}}=-A_{CP}^{D,\text{SM}}=A_L.
\eeq
In each case, the signs follow from analyzing the quark content: for $K_L\to \pi^- \ell^+\nu_\ell$ and $\tau^+\to\pi^+ K_S\bar\nu_\tau$ the neutral kaon is produced as $K^0=\bar s d$, while $D^+\to\pi^+ K_S$ requires $\bar K^0=\bar d s$ (in this case, however, there are corrections from the Cabibbo-suppressed decay mode~\cite{Yu:2017oky}). Indeed, for the $D$-meson decay the corresponding prediction $A_{CP}^{D,\text{SM}}=-3.32(6)\times 10^{-3}$ agrees well with the experimental result~\eqref{D_CP}.
In contrast, while earlier searches had not found evidence for $CP$ violation~\cite{Bonvicini:2001xz,Bischofberger:2011pw}, the latest result by the BaBar collaboration~\cite{BABAR:2011aa}
\beq
\label{CP_tau_exp}
A_{CP}^{\tau,\text{exp}}=-3.6(2.3)(1.1)\times 10^{-3}
\eeq
revealed a striking disagreement with the SM prediction. 
As pointed out in~\cite{Grossman:2011zk}, since the intermediate $K_S$ is reconstructed in terms of a final-state $\pi^+\pi^-$ pair with invariant mass around $\mk$ and a decay time consistent with the $K_S$ lifetime, the prediction~\eqref{CP_SM} might be altered by the exact experimental conditions.  However, the corresponding shift to $A_{CP}^{\tau,\text{SM}}=3.6(1)\times 10^{-3}$ even slightly increases the discrepancy to $2.8\sigma$~\cite{BABAR:2011aa}.

Optimistically, this tension could be considered a hint for BSM physics and it is natural to ask in a first step whether it is possible to account for the difference with non-standard interactions. 
In general, for producing a non-vanishing $CP$ asymmetry one needs the interference of two amplitudes 
\beq
\mathcal{A}_j=|\mathcal{A}_j|e^{i\delta^\text{s}_j}e^{i\delta^\text{w}_j},\qquad j\in\{1,2\},
\eeq
with relative strong and weak phases $\delta^\text{s}=\delta^\text{s}_1-\delta^\text{s}_2$ and $\delta^\text{w}=\delta^\text{w}_1-\delta^\text{w}_2$. Both phases have to be non-vanishing, i.e.\
\begin{align}
A_{CP}&\propto|\mathcal{A}_1+\mathcal{A}_2|^2-|\bar{\mathcal{A}}_1+\bar{\mathcal{A}}_2|^2\notag\\
&=-4|\mathcal{A}_1||\mathcal{A}_2|\sin \delta^\text{s}\sin\delta^\text{w}.
\end{align}
Here the $\bar{\mathcal{A}}_j$ denote the amplitudes with opposite weak phase.
As argued in~\cite{Devi:2013gya,Dhargyal:2016kwp,Dhargyal:2016jgo}, due to the lack of a strong phase, this excludes an explanation using scalar operators, but new tensor interactions were found to be admissible.  
In this paper we will demonstrate that this conclusion relies on erroneous assumptions for the $\pi K$ tensor form factor, provide the corrected expression of the $CP$ asymmetry in terms of the tensor Wilson coefficient, and study the consequences for a possible BSM explanation of~\eqref{CP_tau_exp}.

\section{Kinematics and conventions}

We define momenta according to
\beq
\tau(p_\tau)\to K_S (\pk) + \pi (\ppi) + \nu_\tau (\pnu),
\eeq
with invariant mass $s=(\pk+\ppi)^2$ of the $\pi K$ system. In the following, we are only interested in the singly-differential decay rate $\diff\Gamma/\diff s$, which is obtained after integrating over the remaining angular dependence of the three-body phase space.

For effective operators and form factors we largely follow the conventions of~\cite{Antonelli:2008jg}.  
Due to parity conservation in the $K\to\pi$ matrix elements it suffices to consider the effective Lagrangian
\begin{align}
\label{Lagr_Delta_S1}
\Lagr^{\Delta S=1}_{su}&= -\frac{G_F}{\sqrt{2}} V_{us}
\Big[c_V (\bar s \gamma^\mu u)(\bar \nu \gamma_\mu  \ell)+ c_A (\bar s \gamma^\mu u)(\bar \nu \gamma_\mu\gamma_5  \ell)\notag\\
&+c_S (\bar s u)(\bar \nu \ell) +ic_P (\bar s u)(\bar \nu \gamma_5\ell)\notag\\
&+c_{T} (\bar s \sigma^{\mu\nu}u)(\bar \nu \sigma_{\mu\nu}(1+\gamma_5)\ell)
\Big] + \text{h.c.},
\end{align}
where the Fermi constant $G_F$ and the CKM element $V_{us}$ have been factored out, and 
we have ignored all operators that vanish in the absence of right-handed neutrinos.
The Wilson coefficients are defined at the weak scale in such a way that in the SM
$c_{V}(M_W)=-c_{A}(M_W)=1$ and all others equal to zero. 
The hadronic matrix elements are parameterized via form factors
\begin{align}
 \langle\bar K^0(\pk)\pi^-(\ppi)|\bar s\gamma^\mu u|0\rangle&=(\pk-\ppi)^\mu f_+(s)\notag\\
 &+(\pk+\ppi)^\mu f_-(s),\notag\\
 \langle\bar K^0(\pk)\pi^-(\ppi)|\bar s u|0\rangle&=\frac{\mk^2-\mpi^2}{m_s-m_u}f_0(s),\notag\\
 \langle\bar K^0(\pk)\pi^-(\ppi)|\bar s\sigma^{\mu\nu} u|0\rangle&=i\frac{\pk^\mu\ppi^\nu-\pk^\nu\ppi^\mu}{\mk}B_T(s),
\end{align}
where
\beq
f_-(s)=\frac{\mk^2-\mpi^2}{s}\big(f_0(s)-f_+(s)\big).
\eeq
Taking everything together, we obtain for the differential decay width for $\tau^-\to K_S\pi^-\nu_\tau$ (see also~\cite{Kuhn:1992nz,Finkemeier:1996dh,Devi:2013gya})
\begin{align}
\label{decay_width}
\frac{\diff\Gamma}{\diff s}&=G_F^2|V_{us}|^2S_\text{EW}\frac{\lambda^{1/2}_{\pi K}(s)(\mtau^2-s)^2(\mk^2-\mpi^2)^2}{1024\pi^3\mtau s^3}\notag\\
&\times\bigg[\xi(s)\bigg(|V(s)|^2+|A(s)|^2+\frac{4(\mtau^2-s)^2}{9s\mtau^2}|T(s)|^2\bigg)\notag\\
&\qquad+|S(s)|^2+|P(s)|^2\bigg],
\end{align}
where $\lambda_{\pi K}(s)=\lambda(s,\mpi^2,\mk^2)$, $\lambda(a,b,c)=a^2+b^2+c^2-2(ab+ac+bc)$,
\beq
\xi(s)=\frac{(\mtau^2+2s)\lambda_{\pi K}(s)}{3\mtau^2(\mk^2-\mpi^2)^2},
\eeq
$S_\text{EW}=1.0194$~\cite{Marciano:1985pd,Marciano:1988vm,Braaten:1990ef} encodes the electroweak running down to $\mtau$,
and
\begin{align}
 V(s)&=f_+(s) c_V-T(s),\qquad A(s)=f_+(s) c_A+T(s),\notag\\
 S(s)&=f_0(s)\bigg(c_V+\frac{s}{\mtau(m_s-m_u)}c_S\bigg),\notag\\
 P(s)&=f_0(s)\bigg(c_A-i\frac{s}{\mtau(m_s-m_u)}c_P\bigg),\notag\\
 T(s)&=\frac{3s}{\mtau^2+2s}\frac{\mtau}{\mk}c_T B_T(s).
\end{align}
In the SM case $c_V=-c_A=1$ (and $c_S=c_P=c_T=0$) this reduces to
\begin{align}
\label{decay_width_SM}
 \frac{\diff\Gamma}{\diff s}\bigg|_{\text{SM}}&=G_F^2|V_{us}|^2S_\text{EW}\frac{\lambda^{1/2}_{\pi K}(s)(\mtau^2-s)^2(\mk^2-\mpi^2)^2}{512\pi^3\mtau s^3}\notag\\
 &\times\bigg[\xi(s)\big|f_+(s)\big|^2+\big|f_0(s)\big|^2\bigg].
\end{align}
The general decomposition of the decay width~\eqref{decay_width} already shows why the scalar--vector (pseudoscalar--axial-vector) interference encoded in $S(s)$ ($P(s)$) cannot produce a $CP$ asymmetry: the hadronic form factor $f_0(s)$ factorizes, so that the relative strong phase vanishes. 
This leaves the interference with the tensor operator in $V(s)$ and $A(s)$
as the only possible source for a strong phase.\footnote{The interference  of vector and scalar operator could still contribute to the $CP$ asymmetry due to 
long-distance QED corrections~\cite{Antonelli:2013usa}. 
We  estimate  $|A_{CP}^{\tau,\text{BSM}}|\lesssim 10^{-4} |\Im c_S|$, which is strongly suppressed 
due to the kinematic factor $\xi(s)$, see~\eqref{decay_width_SM}, 
the suppression of $f_0(s)$ compared to $f_+(s)$, 
and  the QED  factor $\Order(\alpha/\pi)$.
The branching ratio for $\tau\to K_S\pi\nu_\tau$ itself already excludes $|\Im c_S|\gtrsim 1$, 
so that even without further 
input the scalar contribution to $A_{CP}^\tau$ is of little phenomenological relevance.
}

Upon neglecting direct $CP$ violation in the SM, the total $CP$ asymmetry can be written as~\cite{Devi:2013gya}
\beq
A_{CP}^\tau=\frac{A_{CP}^{\tau,\text{BSM}}+A_{CP}^{\tau,\text{SM}}}{1+A_{CP}^{\tau,\text{BSM}}A_{CP}^{\tau,\text{SM}}},
\eeq
where
\begin{align}
\label{CP_BSM}
 A_{CP}^{\tau,\text{BSM}}&=\frac{\sin\delta_T^\text{w} |c_T|}{\Gamma_\tau\text{BR}(\tau\to K_S\pi\nu_\tau)}\\
 &\hspace{-1cm}\times\int_{s_{\pi K}}^{\mtau^2}\diff s'\kappa(s')|f_+(s')||B_T(s')|
 \sin\big(\delta_+(s')-\delta_T(s')\big),\notag
\end{align}
$s_{\pi K}=(\mpi+\mk)^2$, and
\beq
\kappa(s)=G_F^2|V_{us}|^2S_\text{EW}\frac{\lambda_{\pi K}^{3/2}(s)(\mtau^2-s)^2}{256\pi^3\mtau^2\mk s^2}.
\eeq
Moreover, $\delta_T^\text{w}$ denotes the phase of $c_T$ relative to $c_V=-c_A=1$, and $\delta_+(s)$, $\delta_T(s)$ are the phases of $f_+(s)$ and $B_T(s)$.

\section{Hadronic form factors}

In~\cite{Devi:2013gya} it was assumed that $B_T(s)$ is constant in such a way that only the phase of $f_+(s)$ remains and produces a sizable $CP$ asymmetry via ~\eqref{CP_BSM}. That this assumption is incorrect can be argued in several ways.  In the context of a vector-meson-dominance picture, the form factor $f_+(s)$ is dominated by the isospin-$I=1/2$, spin-$1$ resonances $K^*(892)$ and the $K^*(1410)$, Breit--Wigner (BW) approximations of which are indeed used to parameterize the experimental decay width for the $\tau\to K_S\pi\nu_\tau$ process~\cite{Epifanov:2007rf}.
However, spin-$1$ resonances can be described equivalently by vector or antisymmetric tensor fields~\cite{Ecker:1988te,Ecker:1989yg}, so that the same resonances that contribute to $f_+(s)$ will appear in $B_T(s)$ as well, most notably the $K^*(892)$.

Beyond the model approach, this conclusion can be derived by analyzing the unitarity relation for the form factors. For the vector current, such constraints from dispersion relations are frequently used to derive a parameterization of the form factor with good analytic properties~\cite{Moussallam:2007qc,Boito:2008fq,Boito:2010me,Bernard:2011ae,Antonelli:2013usa}. In particular, $\pi K$ intermediate states generate an imaginary part according to
\beq
\label{unitarity}
\Im f_+(s)=\frac{\lambda_{\pi K}^{1/2}(s)}{s}f_+(s) \big(f_1^{1/2}(s)\big)^*\theta(s-s_{\pi K}),
\eeq
where the $\pi K$ partial waves $f_l^I(s)$ (with angular momentum $l$) obey the elastic unitarity relation
\beq
\Im f_l^I(s)=\frac{\lambda^{1/2}_{\pi K}(s)}{s}\big|f_l^I(s)\big|^2\theta(s-s_{\pi K}),
\eeq
and can thus be parameterized in terms of the $\pi K$ phase shifts $\delta_l^I(s)$ 
\beq
f_l^I(s)=\frac{s}{\lambda^{1/2}_{\pi K}(s)}e^{i\delta_l^I(s)}\sin \delta_l^I(s).
\eeq
The unitarity relation for the form factor~\eqref{unitarity} then implies that, in the elastic region, the phase of $f_+(s)$ has to coincide with $\delta_1^{1/2}(s)$, a manifestation of Watson's final-state theorem~\cite{Watson:1954uc}. In this way, $\delta_1^{1/2}(s)$ can be considered a model-independent implementation of the $K^*(892)$, which indeed decays almost exclusively to the $K\pi$ channel.

In fact, the tensor form factor obeys the exact same unitarity relation
\beq
\label{ImBT}
\Im B_T(s)=\frac{\lambda^{1/2}_{\pi K}(s)}{s}B_T(s) \big(f_1^{1/2}(s)\big)^*\theta(s-s_{\pi K}),
\eeq
which simply reflects the fact that the $K^*(892)$ is equally well described by a vector or an antisymmetric tensor field. 
The relation~\eqref{ImBT} can be derived explicitly from the $\pi K$ loop integral using standard Cutkosky rules. 
To actually construct a parametrization for $B_T(s)$ this relation is not sufficient because it does not determine the normalization. However, it shows that as long as $\pi K$ states dominate the unitarity relation, the phases of $f_+(s)$ and $B_T(s)$ are identical, so that the corresponding $CP$ asymmetry vanishes.
This statement is exact as long as inelastic states, most notably $\pi\pi K$, are negligible. The next resonance, $K^*(1410)$ decays predominantly via $K^*(1410)\to K^*(892)\pi \to K\pi\pi$, and this indeed requires inelastic contributions that will result in a non-vanishing $CP$ asymmetry. Given the dominance of the $K^*(892)$ resonance, the cancellation in the elastic region will strongly suppress the amount of $CP$ asymmetry that a tensor operator can produce.

\begin{figure}[t]
 \includegraphics[width=\linewidth,clip]{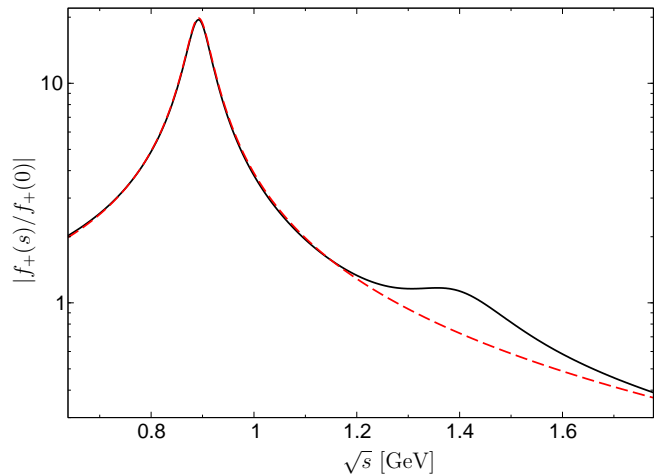}
 \caption{$|f_+(s)/f_+(0)|$ from~\cite{Epifanov:2007rf} (black solid line) in comparison to the Omn\`es factor~\eqref{Omnes} (red dashed line).}
 \label{fig:modulus}
\end{figure}

Empirically, information on the form factor $f_+(s)$ can be derived from the $\tau\to\pi K_S\nu_\tau$ spectrum~\cite{Epifanov:2007rf}, which is strongly dominated by the $K^*(892)$ resonance. For the modulus of the form factors in~\eqref{CP_BSM} we can therefore ignore any inelastic corrections and use the elastic solution of the unitarity relation
\beq
f_+(s)=f_+(0)\Omega(s),\qquad B_T(s)=B_T(0)\Omega(s),
\eeq
in terms of the Omn\`es factor~\cite{Omnes:1958hv} 
\beq
\label{Omnes}
\Omega(s)=\exp\bigg\{\frac{s}{\pi}\int_{s_{\pi K}}^\infty\frac{\delta(s')}{s'(s'-s)}\bigg\}.
\eeq
The phase shift $\delta(s)$ can be identified with $\delta_1^{1/2}(s)$, and be approximated by a BW phase with parameters as determined in~\cite{Epifanov:2007rf}. The resulting modulus is virtually indistinguishable from the experimental fit below the $K^*(1410)$ resonance, see Fig.~\ref{fig:modulus}, and the relative size of the two resonance peaks serves as an indication for the size of the inelastic effects.

\begin{figure}[t]
 \includegraphics[width=\linewidth,clip]{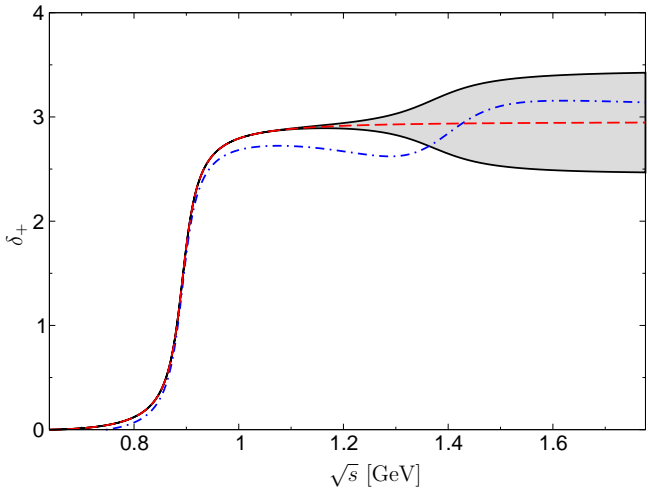}
 \caption{$\delta_+$ from a BW approximation for the $K^*(892)$ (red dashed line) in comparison to the phase from the experimental fit~\cite{Epifanov:2007rf} (blue dot-dashed line). The band represents our estimate of inelastic effects, see main text for details.}
 \label{fig:phase}
\end{figure}

The phase $\delta_+(s)$ cannot be directly taken from experiment, which is only sensitive to the modulus, and its extraction requires the use of a fit function that preserves the analytic structure of the form factor. This is not the case for the fit function used in~\cite{Epifanov:2007rf} (a superposition of BW functions with complex coefficients), see Fig.~\ref{fig:phase}, and indeed the corresponding phase cannot be physical because it does not vanish at threshold and violates Watson's theorem long before the $K^*(1410)$ can possibly have an effect.
Still,  the deviation between the  phase  found to be compatible with the spectrum~\cite{Epifanov:2007rf}  (blue dot-dashed line in Fig.~\ref{fig:phase}), and the elastic phase (red dashed line in Fig.~\ref{fig:phase}) provides a useful indication of the size of inelastic effects. As a simple estimate of the inelastic contribution $\delta^\text{inel}_+(s)$ to $\delta_+(s)$  we add the BW phase for $K^*(1410)\to K^*(892)\pi$ with a coefficient that allows for 
a similar phase motion in the vicinity of the $K^*(1410)$, to arrive at the band shown in Fig.~\ref{fig:phase} (consistent with more refined estimates along the lines of~\cite{Moussallam:2007qc,Boito:2008fq,Boito:2010me,Bernard:2011ae,Antonelli:2013usa}).  

Assuming that inelastic contributions in $\delta_T(s)$ are of similar size (but potentially opposite in sign),   
we take $\delta_+(s) - \delta_T(s) \sim 2 \delta_+^\text{inel}(s)$.
With this at hand,  using $\text{BR}(\tau\to K_S\pi\nu_\tau)=4.04(13)\times 10^{-3}$~\cite{Epifanov:2007rf}, 
$B_T(0)/f_+(0)=0.676(27)$ from lattice QCD~\cite{Baum:2011rm} (see~\cite{Becirevic:2000zi} for an earlier calculation),
and $f_+(0)|V_{us}|=0.2165(4)$ as well as particle masses and couplings from~\cite{Patrignani:2016xqp} (see also~\cite{Antonelli:2009ws}), we estimate  for the $CP$ asymmetry~\eqref{CP_BSM}
\beq
\label{ACP_cT}
\big|A_{CP}^{\tau,\text{BSM}}\big|\lesssim 0.03 |\Im c_T|,
\eeq
about two orders of magnitude less than for the maximum hadronic phase assumed in~\cite{Devi:2013gya}.

\section{Limits on $\boldsymbol{\Im c_T}$}

To further appraise~\eqref{ACP_cT} we now turn to phenomenological constraints on $\Im c_T$. By exploiting $SU(2)$ invariance of the weak interactions very strong limits follow from the
electric dipole moment (EDM) of the neutron and $D$--$\bar{D}$ mixing, as we will demonstrate in the following.

\begin{figure}[t]
 \includegraphics[width=0.6\linewidth,clip]{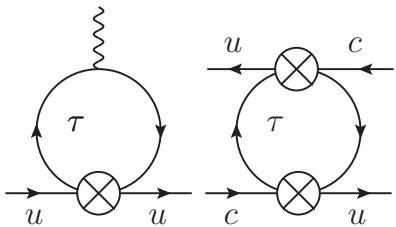}
 \caption{Diagrammatic representation of the electromagnetic dipole operator contributing to the neutron EDM produced by inserting the $(\bar{\tau} \sigma_{\mu \nu}R  \tau)(\bar{u}  \sigma^{\mu \nu} R u)$ operator (left), and the contribution to $D$--$\bar D$ mixing originating from the double insertion of the operator $(\bar{\tau} \sigma_{\mu \nu}R  \tau)(\bar{c}  \sigma^{\mu \nu} R u)$ (right, the second permutation is omitted).}
 \label{fig:RG}
\end{figure}

At a high scale $\Lambda \gg v$, where $v=246\GeV$ is the vacuum expectation value of the Higgs field, the tensor operator contributing to $\tau \to K_S \pi \nu_\tau$ arises from the following $SU(3) \times SU(2) \times U(1)$ gauge-invariant Lagrangian
\beq
\label{eq:LT0}
\Lagr_T  =   C_{abcd}   \ \bar{L}_{La}^i  \sigma_{\mu \nu}  e_{Rb} \, \epsilon^{ij}  \, \bar{q}_{L c}^j  \sigma^{\mu \nu}  u_{Rd}  \ + \ \text{h.c.}, 
\eeq
where $L_L$  and $q_L$ denote the  lepton and quark $SU(2)_L$ doublets,  $e_R$ and $u_R$ are the charged lepton and up-quark  $SU(2)_L$ singlets, 
$i,j$ are $SU(2)_L$ indices, and $a,b,c,d$  
are generation indices.\footnote{In the notation of~\cite{Buchmuller:1985jz,Grzadkowski:2010es} this is the operator $Q^{(3)}_{\ell e qu}$.}  
The tensor operator in~\eqref{Lagr_Delta_S1} is generated from 
\begin{align}
\label{eq:LT1}
\Lagr_T  &= C_{3321}\Big[ 
(\bar{\nu}_{\tau} \sigma_{\mu \nu}R  \tau)(\bar{s}  \sigma^{\mu \nu} R u)\notag\\
&- V_{us} (\bar{\tau} \sigma_{\mu \nu}R  \tau)(\bar{u}  \sigma^{\mu \nu} R u)
\Big]   \ + \ \text{h.c.},
\end{align}
where $R=(1+\gamma_5)/2$, in the second line
terms involving the charm and top quark have been neglected, and 
the Wilson coefficient $C_{3321}$ is related to $c_T$ by
\beq
\label{C3321}
C_{3321} = - \sqrt{2} G_F  V_{us} c_T =  - V_{us} \frac{c_T}{v^2}.
\eeq
In this way, $SU(2)$ symmetry relates the tensor operator relevant for $\tau \to K_S \pi \nu_\tau$ to a neutral current operator involving the $\tau$ and the up quark only. 
The renormalization group evolution~\cite{Jenkins:2013wua} of this operator then produces an up-quark EDM $d_u (\mu)$,
\beq
\Lagr_\text{D}=- \frac{i}{2}  d_u (\mu)  \bar{u} \sigma^{\mu \nu} \gamma_5   u    F_{\mu \nu},
\eeq
via the diagram shown in Fig.~\ref{fig:RG}. Solving the RG following~\cite{Bellucci:1981bs,Buchalla:1989we,Cirigliano:2017azj}
we find
\begin{align}
d_u (\mu) &=   \frac{e \mtau}{v^2}    \frac{V_{us}^2}{\pi^2}   \   \Im c_T (\mu)   \ \log \frac{\Lambda}{\mu}
 \notag\\
&\simeq 3.0 \times \Im c_T (\mu)   \log \frac{\Lambda}{\mu}  \times 10^{-21}  \, e \,\text{cm}.  
\end{align}
Using  the $90\%$ C.L.\ bound $d_n =  g_T^{u} (\mu)  d_u (\mu) <  2.9 \times 10^{-26}\,e\,\text{cm}$~\cite{Baker:2006ts,Afach:2015sja}
  and the recent lattice result~\cite{Bhattacharya:2015esa}
$g_T^{u} (\mu = 2\GeV)  = - 0.233(28)$   we obtain   ($\mu_\tau = 2\GeV$)
\beq
| \Im c_T (\mu_\tau) |   \leq  \frac{4.4 \times 10^{-5}}{ \log \frac{\Lambda}{\mu_\tau}}
\lesssim  10^{-5},
\label{eq:cTbound}
\eeq
where the last inequality holds for $\Lambda\gtrsim 100\GeV$.
This bound is based on the assumption that there are no other contributions to  the neutron EDM 
canceling the effect of $c_T$. However, for values of $\Im c_T (\mu_\tau)  \sim 0.1$ required to explain the tau $CP$ asymmetry, the $c_T$ contribution alone would predict a neutron EDM four orders of magnitude larger than the current bound,
requiring an extraordinary cancellation at the level of one part in $10^4$. 

\begin{figure}[t]
 \includegraphics[width=\linewidth,clip]{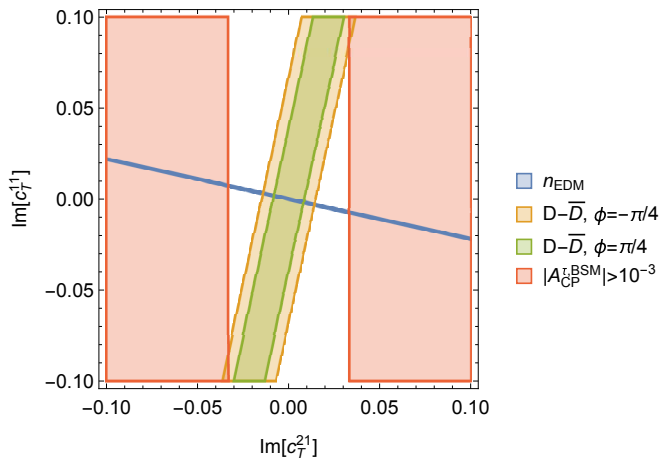}
 \caption{Allowed regions in the $\Im c_T^{21}$--$\Im c_T^{11}$ plane from the neutron EDM and $D$--$\bar D$ mixing (for $\phi=\pm \pi/4$ and $\Lambda=1\,\text{TeV}$), compared to the favored region from the $\tau \to K_S \pi \nu_\tau$ $CP$ asymmetry. The exclusion regions for $\phi=\pm \pi/4$ differ due to the asymmetric form of the fit result in~\cite{Bevan:2014tha}.}
 \label{fig:exclusion}
\end{figure}

Such a cancellation could in principle occur with operators related to the flavor structure $C_{3311}$ in~\eqref{eq:LT0}, since the neutron EDM is sensitive to the combination $V_{ud}\Im c_T^{11}+V_{us}\Im c_T^{21}$, where $c_T^{21}=c_T$ and $c_T^{11}$ is defined analogously to~\eqref{C3321}.
However, yet another combination appears in $D$--$\bar D$ mixing, which is very sensitive to the imaginary part of the Wilson coefficients (as for example defined in~\cite{Bona:2007vi})
\beq
\label{Wilson_D_mixing}
C_2'=\frac{1}{2}C_3'=4G_F^2\frac{\mtau^2}{\pi^2}\log\frac{\Lambda}{\mu_\tau}V_{us}^2\big(V_{cd}c_T^{11}+V_{cs}c_T^{21}\big)^2,
\eeq
where we have neglected the effect of external momenta, i.e.\ the mass of the charm quark. 
Using the global fit of~\cite{Bevan:2014tha} and assuming the phase of $V_{cd}c_T^{11}+V_{cs}c_T^{21}$ to be equal to $\phi=\pm\pi/4$,\footnote{In general, the constraint is diluted by $\sqrt{|\tan\phi|}$ and therefore disappears for $\phi=\pm\pi/2$.} this leads to the situation depicted in Fig.~\ref{fig:exclusion}. 
Since~\eqref{Wilson_D_mixing} requires the insertion of two effective operators,  
the leading contribution here is of dimension $8$, while in an ultraviolet complete model there is in general already a dimension-$6$ contribution, making the bounds from $D$--$\bar D$ mixing even stronger than the one shown in Fig.~\ref{fig:exclusion}.
To evade all bounds, one would therefore not only have to cancel the $c_T$ contribution to the neutron EDM at the level of $10^{-4}$, but also tune the combination $V_{cd}c_T^{11}+V_{cs}c_T^{21}$ close to purely imaginary to evade the constraint from $D$--$\bar D$ mixing.  

\section{Conclusions}

In this article we examined non-standard  contributions to the $CP$ asymmetry in $\tau\to K_S\pi\nu_\tau$. We find that at the dimension $6$ level only the tensor operator can lead to direct $CP$ violation, with negligible QED corrections from the scalar operator. However, the effect of the tensor operator is much smaller than previously estimated as a consequence of Watson's final-state-interaction theorem. Therefore, a very large imaginary part of the Wilson coefficient of the tensor operator would be required in order to account for the current tension between theory and experiment. In fact, we find in a model-independent analysis that this is in general in conflict with the bounds from the neutron EDM and $D$--$\bar D$ mixing, making a BSM explanation (realized above the electroweak breaking scale) highly improbable.

Nonetheless, a confirmation of the current BABAR measurement by Belle and/or Belle II would have intriguing consequences. In the absence of fine tuning, it would point towards the existence of light BSM physics (realized below the electroweak breaking scale) so that our model-independent bounds could be evaded. We hope that the present analysis provides additional motivation to pursue such a measurement.

\section*{Acknowledgments}
\begin{acknowledgments}
This research is  supported by the U.S.\ Department of Energy,
 Office of Science, Office of Nuclear Physics, under contracts
DE-AC52-06NA25396 and DE-FG02-00ER41132. 
A.C.\ is supported by an
Ambizione Grant of the Swiss National Science Foundation (PZ00P2\_154834).
\end{acknowledgments}

\end{document}